\documentclass[twocolumn,showpacs,amsmath,amssymb]{revtex4}
\usepackage{graphicx}
\usepackage{dcolumn}
\usepackage{bm}
\usepackage[usenames,dvipsnames]{color} 
\usepackage[normalem]{ulem} 
\newcommand{\beq}{\begin{equation}}
\newcommand{\eeq}{\end{equation}}
\newcommand{\bea}{\begin{eqnarray}}
\newcommand{\eea}{\end{eqnarray}}
\newcommand{\eps}{\varepsilon}

\newcommand{\mkrm}[1]{}           

\begin{document}

\author{S. V. Tolokonnikov} \affiliation{National Research Centre ``Kurchatov
Institute'', 123182 Moscow, Russia} \affiliation{Moscow Institute of Physics and Technology,
Dolgoprudny}

\author{I. N. Borzov}
\affiliation{National Research Centre ``Kurchatov Institute'', 123182 Moscow, Russia}
\affiliation{Joint Institute for Nuclear Research, 141980 Dubna, Russia}

\author{M. Kortelainen}
\affiliation{Department of Physics, P.O. Box 35 (YFL), University of Jyvaskyla, FI-40014
Jyvaskyla, Finland}
\affiliation{Helsinki Institute of Physics, P.O. Box 64, FI-00014 University of Helsinki, Finland}

\author{Yu. S. Lutostansky}
\affiliation{National Research Centre ``Kurchatov Institute'', 123182 Moscow, Russia}

\author{E. E. Saperstein}
\affiliation{National Research Centre ``Kurchatov Institute'', 123182 Moscow, Russia}
\affiliation{National Research Nuclear University MEPhI, 115409 Moscow, Russia}

\title{Alpha-decay energies of superheavy nuclei for the Fayans functional}


\begin{abstract}
Alpha-decay energies for several chains of super-heavy nuclei are calculated within the
self-consistent mean-field approach by using the Fayans functional FaNDF$^0$. They are compared to the
experimental data and predictions of two Skyrme functionals, SLy4 and SkM*, and of the macro-micro
method as well. The corresponding lifetimes are calculated with the use of the semi-phenomenological
formulas by Parkhomenko and Sobiczewski and by Royer and Zhang.
\end{abstract}

\maketitle

\section{Introduction}
During  the last decade, remarkable progress has been achieved in a synthesis of superheavy nuclei
belonging  to the so-called ``stability island'', the elements with charge numbers $Z \simeq 114 -
120$, predicted about fifty years ago \cite{Myers-Swiat,Strut}. Two isotopes of the element with
$Z=116$, named later Lv, and one with $Z=118$ were created first by Oganessian {\it et. al.}
\cite{Oganes_2006}. After the element $118$ two  isotopes of the element $Z=117$ were synthesized
\cite{Oganes_2010}. The latter element was investigated in more details in Refs.
\cite{Oganes_2011,Oganes_2012,Oganes_2013,Z117_new}. An attempt \cite{Oganes_2009} to create the
element with $Z=120$ turned out to be unsuccessful, however, evidently, efforts in this direction will
be continued.

In fact, the stability island turned out to be ``the stability shallow'', as all of  these new nuclei
are not stable. Although their lifetimes are large in the ``nuclear scale'', they are usually  not
sufficiently long lived to be detected  in a usual way.  All of them undergo alpha-decay, and the
authors of the experimental works cited above detected, for each primary nucleus, a  chain of several
alpha-decays by measuring the alpha-decay energies $E_{\alpha}$ with high accuracy,  and the
respective lifetimes $T_{\alpha}$, leading to a nucleus which was already known. Therefore, a
theoretical support is desirable to make such kind of indirect identification of the new superheavy
nuclei more reliable.

For even-even nuclei, in which the transition occurs between the $0^+$ ground states, the alpha-decay
energies $E_{\alpha}(Z,N)$ are determined, with allowance for the recoil effect, in terms of the mass
difference between nuclei related by alpha decay: \beq Q_{\alpha}(Z,N)=M(^A_ZX_N) -
M(^{A-4}_{Z-2}Y_{N-2})-M_{\alpha}. \label{EQ}\eeq For odd and odd-odd nuclei, a correction to this
simple formula could appear due to a possible excitation of the parent and/or daughter nucleus which
may occur in the real experimental situation.

In Refs. \cite{Oganes_2006,Oganes_2010,Oganes_2011,Oganes_2012,Oganes_2013}, the experimental data for
$Q_{\alpha}$ were compared with predictions \cite{Muntian-1,Muntian-2} of the so-called macro-micro
method (MMM) \cite{MMM,MMM_1}. In this method, the binding energy of a nucleus is found as the sum of
two terms, a macroscopic energy which  was calculated on the basis of the liquid-drop model and a
shell correction energy which  was found according to the  Strutinsky method \cite{Strut}. Thus, two
independent sets of parameters are used in the MMM, one for the nuclear droplet model and the second
for the Shell Model potential.  In general, such a comparison confirmed the identification of new
superheavy nuclei. Recently, in a comprehensive study of superheavy nuclei \cite{Heenen-2015},
predictions of the MMM, including $\alpha$-decay characteristics, were compared to those from the
nuclear energy density functional (EDF) method.  Two Skyrme EDFs were used, SkM* \cite{skms} and SLy4
\cite{sly4to7},   for which the parametrizations can be also found in the review article
\cite{criteria}.   A relativistic EDF DD-PC1 \cite{DD-PC1} was also used for a comparison. The main
bulk of calculations was carried out within   the self-consistent mean-field (SCMF) approach. In
addition, several examples of beyond mean-field results were presented which explicitly considered
collective correlations related to the symmetry restoration and fluctuations of the collective
coordinates. It was shown that the SCMF described well $\alpha$-decay and $\beta$-decay energies. At
the same time, beyond mean-field corrections are important in order to find the correct deformation
energy curves and fission barrier heights.

The finite-range EDF by Fayans {\it et al.} \cite{Fay1,Fay4,Fay} was used first to find $Q_{\alpha}$
for new superheavy $\alpha$-chains in  Ref. \cite{alpha-13}, within a partially self-consistent
calculation, with an approximate consideration  of the deformation energy.  The total binding energy
of a deformed nucleus was presented as a sum of two terms, $E_B(N,Z)=E_{\rm sph}(N,Z)+ E_{\rm
def}(N,Z)$, where the main, spherical one, was found for a Fayans EDF, whereas the ``deformation''
addendum was taken from published tables for two Skyrme EDFs. To be definite, a version DF3-a
\cite{DF3-a} of the initial Fayans EDF DF3 \cite{Fay4,Fay} was used for the spherical energy. For the
deformation energy, the EDF HFB-17 \cite{HFB-17,site} was used for nuclei with $Z\le 110$  and MSk7
\cite{MSk7}, for $Z > 110$. Accuracy of such semi-self-consistent calculations turned out to be only a
bit worse compared to that of the MMM approach.

The absence of a deformed Fayans EDF solver  was the reason of the use of such non-consistent ansatz
in Ref. \cite{alpha-13}. Recently, however,  such a code has been  developed \cite{Fay-def}. A
localized version FaNDF$^0$ \cite{Fay5} of the general finite range Fayans EDF was used, making the
surface term more similar  to the Skyrme one. This allowed to employ, with some modifications, the
computer code HFBTHO \cite{code}, developed originally for Skyrme EDFs. First applications of this
code to deformed nuclei \cite{Fay-def,Fay-def1,drip-2n,Pb-def} with the use of the original set of
parameters of the EDF FaNDF$^0$ \cite{Fay5} found for spherical nuclei turned out to be rather
successful.  The goal of the present work is to provide a  new, completely self-consistent,
calculation of $Q_{\alpha}$ energies  for six superheavy $\alpha$-chains with the Fayans EDF. For a
comparison, calculations with two Skyrme EDFs, SLy4 and SkM*, are carried out. These two
parametrizations are rather commonly employed in various Skyrme EDF calculations.   In particular,
they were used as representatives of non-relativistic EDFs in \cite{Heenen-2015}.   Thus, this part of
our calculations repeats that of \cite{Heenen-2015} and partly earlier study of~\cite{(Cwi99)}. With
these two Skyrme parametrizations, when considering low-lying collective excitations in spherical
nuclei, the older SkM* was found more successful than SLy4~\cite{BE2-SHF}. Predictions of the
self-consistent methods are compared to the MMM of Refs. \cite{Muntian-1,Muntian-2}.

Theoretical predictions for $Q_{\alpha}$ are  important not only by  itself but also to find  the
lifetime $T_{\alpha}$. Indeed, the latter is governed mainly by the exponential Gamow factor
\cite{Gamow}, which is found almost unambiguously in terms of $Q_{\alpha}$ by finding the
penetrability of the Coulomb barrier in the daughter nucleus for an emitted alpha particle.
Unfortunately, at the present there is no a reliable microscopic theory for the pre-exponential factor
as this is a very complicated many-body problem. It is worth mentioning recent studies in this
direction \cite{Peltonen,Betan,micr1,Ward}. Some of them are rather promising but do not provide  a
simple tool for systematic calculations. In such a situation, the phenomenological approaches are more
practical. Most of them are close to the classical Viola-Seaborg formula \cite{VS}, which involves
seven phenomenological parameters. Here, just as in \cite{alpha-13}, we use the five-parameter
modification of this formula \cite{Par-Sobich} by Parkhomenko and Sobiczewski (PS). For a comparison,
similar calculations were repeated  with the use of more recent 12-parameter formula by Royer and
Zhang (RZ) \cite{Royer}.

\section{ Fayans EDF predictions for $Q_{\alpha}$ values in superheavy nuclei}

In this section, we give  the SCMF  predictions of the Fayans EDF FaNDF$^0$ \cite{Fay5} for six
$\alpha$-decay chains which begin from the following parent superheavy nuclei: $^{294}118$,
$^{294}117$, $^{293}117$, $^{291}$Lv, $^{288}115$, and $^{287}115$. For completeness, we write down
explicitly the main ingredients of this EDF. In the Fayans method, the ground state energy of a
nucleus is considered as a functional of normal density  $\rho$ and anomalous density $\nu$, as  \beq
E_0=\int {\cal E}[\rho({\bf r}),\nu({\bf r})] d^3r,\label{E0} \eeq  where the isotopic indices and the
spin-orbit densities are omitted for  brevity.

The volume part of the EDF, ${\cal E}^{\rm v}(\rho)$, is taken  as a fractional function of densities
$\rho_+=\rho_n+\rho_p$ and $\rho_-=\rho_n-\rho_p$: \beq {\cal E}^{\rm v}(\rho)=C_0 \left[ a^{\rm
v}_+\frac{\rho_+^2}4 f^{\rm v}_+(x) + a^{\rm v}_-\frac{\rho_-^2}4 f^{\rm v}_-(x)\right], \label{EDF_v}
\eeq where \beq f^{\rm v}_+(x)=\frac{1-h^{\rm v}_{1+}x^{\sigma}}{1+h^{\rm v}_{2+}x^{\sigma}}
\label{fx_vpl} \eeq and \beq f^{\rm v}_-(x)=\frac{1-h^{\rm v}_{1-}x}{1+h^{\rm v}_{2-}x}.
\label{fx_vmi} \eeq Here, $x=\rho_+/\rho_0$ is the dimensionless nuclear density, $\rho_0=2(k_{\rm
F}^{0})^3/3\pi^2$  being the equilibrium symmetric nuclear matter density. The coefficient
$C_0=(dn/d\eps_{\rm F})^{-1}=\pi^2/(k_{\rm F}^{0})m)$  is the usual  normalization factor used in the
theory of finite Fermi system (TFFS) \cite{AB1}, the inverse density of states at the Fermi surface.
The power parameter $\sigma=1/3$ is chosen in the FaNDF$^0$ functional, in contrast to the case for
DF3 or DF3-a, where $\sigma=1$ is used. The dimensionless  parameters in Eqs.
\ref{EDF_v}--\ref{fx_vmi} are the same as  in \cite{Fay5}: $a^{\rm v}_+=-9.559$, $h^{\rm
v}_{1+}=0.633$, $h^{\rm v}_{2+}=0.131$, $a^{\rm v}_-=4.428$, $h^{\rm v}_{1-}=0.25$, and $h^{\rm
v}_{2-}=1.300$. They correspond to the following characteristics of nuclear matter: the equilibrium
density $\rho_0=0.160$ fm$^{-3}$ (the corresponding mean radius parameter is $r_0=1.143$ fm), the
energy per particle $\mu=-16.0$ MeV, the  incompressibility  $K_0=220$ MeV, and the asymmetry energy
coefficient  of  $a_{\rm sym}=30.0$ MeV.   Higher derivatives of the EDF of nuclear matter over the
densities $\rho_+, \rho_-$, suggested in \cite{criteria} for the Skyrme EDFs as a set of criteria for
functionals, are given in \cite{Fay-def} for the FaNDF$^0$ functional.

 If one sets  $h^{\rm v}_{2+}=h^{\rm v}_{2-}=0$, the volume part of the
Fayans EDF reduces to the form typical for Skyrme EDFs. As was discussed in detail in \cite{Fay-def},
the ``Fayans denominator'' and the use of the bare mass $m^*=m$, both peculiarities of the Fayans EDF
method, are generically related to the self-consistent TFFS \cite{KhS},  reflecting  in a hidden form
the energy-dependence effects inherent to this approach.

Until recently, the Fayans EDF was applied to spherical nuclei only. These applications were rather
successful, in comparison with analogous Skyrme  Hartree-Fock-Bogoliubov (HFB)  calculations. They
included the analysis  of the magnetic \cite{mu1,mu2} and quadrupole \cite{QEPJ,QEPJ-Web} moments in
odd nuclei,  of characteristics of the first $2^+$ excitations in even semi-magic nuclei
\cite{BE2,BE2-Web}, of charge radii \cite{Sap-Tolk},  and of beta-decay \cite{Borz} as well. In
addition, single-particle spectra of seven magic nuclei was  described with high accuracy
\cite{Levels}. A short review comparison of predictions of Fayans and Skyrme EDFs for these phenomena
in spherical nuclei was  given in Ref. \cite{compar-spher},  and more detailed one, in Ref.
\cite{compar-YAF}.

The use of the local approximation for the Yukawa finite-range function, ${\rm Yu}(r)\to 1-r_c^2
\nabla^2$, in the DF3-like EDFs leads to the  following structure of the surface term  of the
FaNDF$^0$ functional: \beq {\cal E}^{\rm s}(\rho)=C_0 \frac 1 4 \frac { a^{\rm s}_+ r_0^2(\nabla
\rho_+)^2}{1+h^{\rm s}_{+}x^\sigma+h^{\rm s}_\nabla r_0^2(\nabla x_+)^2}, \label{EDF_s}\eeq with
$h^{\rm s}_{+}=h^{\rm v}_{2+}$, $a^{\rm s}_+=0.600$, $h^{\rm s}_\nabla=0.440$. This approximation is
the main point which permitted in \cite{Fay-def} to modify the Skyrme HFB computer code HFBTHO
\cite{code} for the Fayans EDF. This code solves the HFB equations in axially symmetric harmonic
oscillator basis by assuming time-reversal symmetry.

 Here, just as in \cite{Fay-def}, we use a two-parameter form for the anomalous
term of the EDF:  \beq {\cal E}_{\rm anom}= C_0 \sum_{i=n,p} \nu_i^{\dag}({\bf r}) f^{\xi}(x_+({\bf
r}))\nu_i({\bf r}), \label{Eanom}\eeq where the density-dependent dimensionless effective pairing
force is \beq f^{\xi}(x_+)= f^{\xi}_{\rm ex} + h^{\xi} x_+. \label{fksi}\eeq Two models for pairing
will be used, the volume one with  $h^{\xi}=0$, and the surface model, with $h^{\xi} \simeq
-f^{\xi}_{\rm ex}$. As we use the zero-range pairing force, a cut-off should be used in solving the
gap equation. Here we use the same cut-off energy $E_{\rm cut}=60$ MeV as in \cite{Fay-def}.
 The standard procedure to solve the HFB equations is used.   The code \cite{code} for Skyrme
EDFs and its analogue for the Fayans EDF used in \cite{Fay-def} and here makes it possible
 to restore particle number approximately within the Lipkin--Nogami scheme. Its effect is
significant for the total pairing energy but negligible (about $0.1$--$0.2\;$MeV) for the binding
energy differences entering to the $Q_{\alpha}$ values.

For odd nuclei, we used the equal filling approximation  for the quasiparticle blocking prescription
when solving the HFB equations. We stress that, within this prescription, the mean-field of the odd
(or odd-odd) nucleus is solved fully self-consistently. The equal filling approximation has been
examined in many works, for example, in Refs.~\cite{odd1,odd2}. In the latter, it was estimated that
this approximation introduces inaccuracy of the order of $0.1$--$0.2\;$MeV for the binding energy.
This is much smaller than the differences between various EDFs for the predicted $Q_{\alpha}$ values.

\begin{figure}
\centerline {\includegraphics [width=0.9\columnwidth]{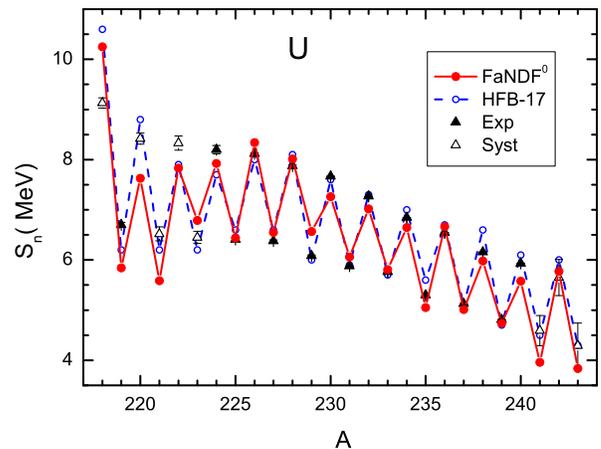}} \caption{The $S_n$ values for the U
isotopic chain.} \label{fig:SnU}
\end{figure}

Pairing interaction influences significantly the one-neutron separation energies \beq
S_n(N,Z)=B(N,Z)-B(N-1,Z). \label{Sn} \eeq  In figure \ref{fig:SnU}, they are shown for uranium
isotopes, calculated with the FaNDF$^0$ EDF, with two models of pairing specified above. The
corresponding values of the parameters of Eq.  (\ref{fksi}) are ($f^{\xi}=-0.440,\, h^{\xi}=0$) for
the volume pairing and ($f^{\xi}=-1.433,\, h^{\xi}=1.375$) for the surface one. In this work, the
calculation scheme for the Fayans EDF is the same as that in Ref.~\cite{Fay-def} for the uranium
chain. In particular, the HFB equations were solved in a basis of 25 oscillator shells. Comparison is
made with experimental data \cite{mass} and predictions from the Skyrme EDF HFB-17 EDF
\cite{HFB-17,site}. We see that the Fayans EDF with both pairing models reproduce experimental data
sufficiently well, with an accuracy comparable with that of the Skyrme EDF HFB-17.

\begin{figure*}
\centerline {\includegraphics [width=0.7\textwidth]{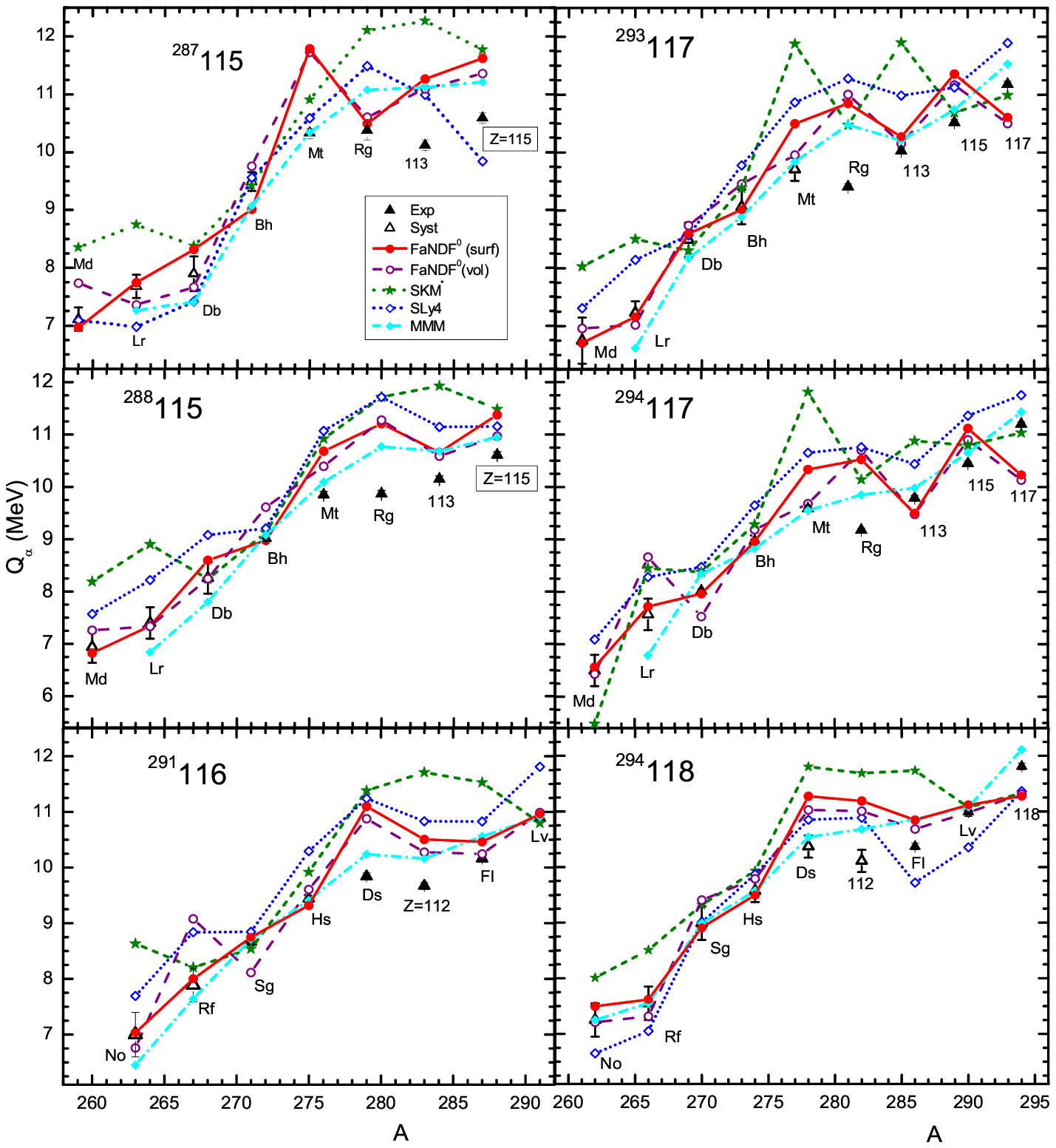}} \caption{The $Q_\alpha$ values for alpha
decay chains starting from nuclei $^{287}115$, $^{293}117$, $^{288}115$, $^{294}117$, $^{291}$Lv, and
$^{294}118$. Theoretical results are shown for FaNDF$^{0}$ with surface and volume pairing, for SkM*
and SLy4 Skyrme EDFs, and for the MMM model. These are compared to experimental and estimated values.}
\label{fig:Qalpha}
\end{figure*}

Next, we investigate  the alpha-decay energies $Q_{\alpha}$. The results of calculations are presented
in figure \ref{fig:Qalpha} and in Table~\ref{table1}. For the Fayans EDF, predictions of the volume
and surface pairing  are given. A comparison is made with the data and predictions of two Skyrme EDFs,
the SLy4 \cite{sly4to7} and SkM* \cite{skms}. Corresponding  results are found by us with the use of
the code \cite{code}. In this case, 20 oscillator shells were  used, in accordance with \cite{msu}.
For both of the Skyrme EDFs,  the mixed  pairing is used \cite{code}: \beq V_{\rm pair}^{n,p}({\bf
r},{\bf r}') = V_0^{n,p}\left(1- \alpha \frac {\rho({\bf r})} {\rho_c} \right) \delta ({\bf r} - {\bf
r}'), \eeq $\alpha=0.5$, with obvious notation.  The paring strength is taken to be  the same for
neutrons and protons, namely $V_0^n=V_0^p = -259.0\;$MeV. For completeness, we included also
predictions of the MMM method taken from the tables in \cite{Muntian-1}, $Z=102$--109, and
\cite{Muntian-2}, $Z=110$--120. Several empty places in the MMM column of Table~\ref{table1} are
caused with absence of the corresponding values in the tables cited above.

A remark should be made about the experimental $Q_\alpha$ values used in the tables and figures 2--7.
They are taken mainly from the references cited in the caption of Table~\ref{table1}, where the
alpha-energies $E_\alpha$ were measured, and  $Q_\alpha$ being found by accounting the recoil effect.
This recipe may be doubtful in the case of odd and odd-odd nuclei where the $\alpha$-transition
between ground states can be forbidden, and the observed transition may connect excited states. The
analysis of the lifetimes $T_\alpha$ may help to clear up this point. They will be analyzed in the
next Section. There are also cases when experimental $Q_\alpha$ values are not available. In this kind
of situation we use estimated masses from mass systematics of Ref.~\cite{mass}. They are shown with
the open triangles in the figures.

\begin{table*}[]
  \caption{$\alpha$-decay energies $Q_{\alpha}$
(MeV) of superheavy nuclei. In each line, $Z$ and $A$ correspond to the parent nucleus.
FaNDF$^0$(surf) and FaNDF$^0$(vol) denote the Fayans EDF with the surface and volume pairing,
correspondingly. Experimental data are taken from \cite{Oganes_2006} ($^{a)}$), \cite{Z117_new}
($^{b)}$), \cite{Oganes_2013} ($^{c)}$), \cite{mass}, compilation of several experiments, ($^{d)}$),
\cite{Oganes_2005} ($^{e)}$). The estimated  data ($^{(\rm syst)}$) are taken from the systematics in
\cite{mass}.} \label{table1}

\begin{tabular}{c c  c r r c l}
\hline
\hline\noalign{\smallskip}

Nuclei&\; FaNDF$^0$(surf) \;&\; FaNDF$^0$(vol) \;&\; SLy4\;&\; SkM*\;&\;MMM \;&\;\;\;\;Exp. \\

\hline
\noalign{\smallskip}
$^{294}{\rm 118}$ & 11.278&  11.310&   11.364&  11.330& 12.11 & 11.81 (0.06)$\,^{a)}$ \\

$^{290}$Lv        & 11.116&  10.981&   10.354&  11.074 & 11.08 & 10.99 (0.08)$\,^{a)}$\\

$^{286}$Fl        &  10.847&  10.683&   9.719&  11.734 & 10.86 & 10.37 (0.06)$\,^{a)}$\\

$^{282}$Cn        &  11.188&  11.005&   10.886& 11.688 & 10.68 &  10.11 (0.2)$\,^{(\rm syst)}$\\

$^{278}$Ds&  11.272&  11.026&  10.849&  11.802  & 10.54 & 10.37 (0.2)$\,^{(\rm syst)}$\\

$^{274}$Hs&   9.497&   9.788&  9.870 &  9.932  &  9.55 & 9.57 (0.2)$\,^{(\rm syst)}$\\

$^{270}$Sg&   8.913&   9.407&   9.004&   9.319 &  8.74 &  8.99 (0.3)$\,^{(\rm syst)}$\\

$^{266}$Rf&   7.617&   7.316&   7.052&   8.508 &  7.05 &  7.55 (0.3)$\,^{(\rm syst)}$\\

$^{262}$No&   7.497&   7.210&   6.649&   7.932 &  6.86 &  7.25 (0.3)$\,^{(\rm syst)}$\\

\hline\noalign{\smallskip}

$^{294}$117&   10.225&   10.131&   11.629&  11.033& 11.43 &11.20  (0.04)$\,^{b)}$ \\

$^{290}$115&   11.117&   10.898 &  10.143&  10.792& 10.65 &10.45  (0.04)$\,^{b)}$ \\

$^{286}$113&    9.494&    9.479&    8.792&  10.879&  9.98 & 9.79  (0.05)$\,^{c)}$ \\

$^{282}$Rg&   10.524&   10.701&   10.037 &  11.512&  9.85 & 9.18  (0.03)$\,^{b)}$\\

$^{278 }$Mt&   10.334&    9.681&    9.851&  10.442&  9.55 & 9.59  (0.03)$\,^{b)}$\\

$^{274 }$Bh&    8.957&    9.187&    8.911&  9.285&  8.83 & 8.97  (0.03)$\,^{b)}$\\

$^{270 }$Db&    7.966&    7.528&    7.643&   8.376&  8.34 & 8.02  (0.03)$\,^{b)}$ \\

$^{266 }$Lr&    7.717&    8.663&    7.993&   8.449&  6.79 & 7.57 (0.3)$\,^{(\rm syst)}$\\

$^{262 }$Md&    6.564&    6.431&    6.011&   7.621&  -    & 6.5 (0.3)$\,^{(\rm syst)}$\\

\hline\noalign{\smallskip}

$^{293}$117&   10.605&   10.502&   10.976& 10.990& 11.53 & 11.18 (0.05)$\,^{d)}$\\

$^{289}$115&   11.356&   11.167&   10.035& 10.688& 10.74 & 10.52 (0.05)$\,^{d)}$\\

$^{285}$113&   10.271&   10.154&    9.592&  11.900& 10.21 & 10.03 (0.05)$\,^{d)}$\\

$^{281 }$Rg&   10.850&   11.003&   10.719&  11.736&  10.48 &  9.41 (0.05)$\,^{c)}$\\

$^{277 }$Mt&   10.498&    9.956&   10.007&  10.618&  9.84 &  9.71 (0.2)$\,^{(\rm syst)}$\\

$^{273}$Bh&    9.014&    9.456&    9.170 &  9.386&  8.89 &  9.06 (0.3)$\,^{(\rm syst)}$\\

$^{269}$Db&   8.596&    8.739&     7.902&   8.305&  8.17 &  8.49 (0.3)$\,^{(\rm syst)}$\\

$^{265 }$Lr&    7.162&    7.021&    7.208&   8.503&  6.62 &  7.23 (0.2)$\,^{(\rm syst)}$\\

$^{261 }$Md&    6.708&    6.960&    6.400&   8.030&  -    &  6.75 (0.4)$\,^{(\rm syst)}$\\

\hline\noalign{\smallskip}

$^{291 }$Lv&   10.959&   10.987&    10.557&  10.798& 10.91 & 10.89 (0.07)$\,^{a)}$\\

$^{287 }$Fl&   10.463&   10.243&     9.130&  11.528&  10.56 & 10.16 (0.06)$\,^{a)}$\\

$^{283 }$Cn&   10.506&   10.275&    10.499&  11.731&  10.16 & 9.67  (0.06)$\,^{a)}$\\

$^{279 }$Ds&   11.097&   10.876&    10.234&  11.361&  10.24 & 9.84  (0.06)$\,^{a)}$\\

$^{275 }$Hs&    9.321&    9.606&    9.600&  9.920&   9.41 & 9.44  (0.06)$\,^{a)}$\\

$^{271 }$Sg&    8.747&    8.115&    8.554&  8.541&   8.71 & 8.67  (0.08)$\,^{a)}$\\

$^{267 }$Rf&    7.998&    9.077&    8.046& 8.204&    -   & 7.89 (0.3)$\,^{(\rm syst)}$\\

$^{263 }$No&    7.031&    6.759&    6.287&  8.634&   6.45 & 7.0 (0.4)$\,^{(\rm syst)}$\\

\hline\noalign{\smallskip}

$^{288}$115&   11.376&   10.968&    9.389&  11.485 & 10.95 & 10.61 (0.06)$\,^{e)}$\\

$^{284}$113&   10.661&   10.588 &  10.393 & 11.930  & 10.68 & 10.15 (0.06)$\,^{e)}$\\

$^{280 }$Rg&   11.208&   11.274 &  11.023 & 11.717  & 10.77 &  9.87 (0.06)$\,^{e)}$\\

$^{276 }$Mt&   10.681&   10.393 &  10.490  &10.919  & 10.09 &  9.85 (0.06)$\,^{e)}$\\

$^{272 }$Bh&    8.978&    9.612 &   8.836 & 9.128  &  9.08 &  9.02 (0.06)$\,^{e)}$\\

$^{268 }$Db&    8.598&    8.245 &   8.514 & 8.239  &  7.90 &  8.26  (0.3)$\,^{e)}$\\

$^{264 }$Lr&    7.343&    7.334 &   6.606 & 8.902  &  6.84 & 7.4 (0.3)$\,^{(\rm syst)}$\\

$^{260 }$Md&    6.823&    7.259 &   6.919 & 8.189  &   -   & 6.94 (0.3)$\,^{(\rm syst)}$\\

\hline\noalign{\smallskip}

$^{287}$115&   11.628&   11.362&   9.840&  11.772&  11.21 & 10.59 (0.09)$\,^{e)}$\\

$^{283}$113&   11.264&   11.090&   10.990& 12.269&  11.12 & 10.12 (0.09)$\,^{e)}$\\

$^{279 }$Rg&   10.494&   10.607&   11.489& 12.107& 11.08 &  10.37 (0.16)$\,^{e)}$\\

$^{275 }$Mt&   11.789&   11.726&   10.585& 10.909& 10.34 & 10.33  (0.09)$\,^{e)}$\\

$^{271 }$Hs&    9.010&    9.760&    9.557&  9.425&   9.07 &  9.49 (0.16)$\,^{(\rm syst)}$\\

$^{267 }$Db&    8.312&    7.667&    7.412&  8.378&   7.41 &   7.9 (0.3)$\,^{(\rm syst)}$\\

$^{263 }$Lr&    7.742&    7.355&    6.979&  8.744&   7.26 &  7.68 (0.2)$\,^{(\rm syst)}$\\

$^{259 }$Md&    6.968&    7.737&    7.091&  8.352 &   -   &  7.11 (0.2)$\,^{(\rm syst)}$\\

 \hline \hline
\end{tabular}
\end{table*}

By eye, the accuracy of the MMM  is higher compared to  self-consistent approaches. Among the latest
EDFs,  the Fayans one, with both the models for pairing, has the accuracy comparable with  both the
Skyrme ones. To estimate it quantitatively, we present in Table~\ref{table2} the differences \beq
\delta Q_{\alpha} = Q_{\alpha}^{\rm theor} - Q_{\alpha}^{\rm exp}, \label{dQalp} \eeq with obvious
notation. In the last column, the  mark ``{(\rm syst)}'' is used for the mass systematics data given
in Ref.~\cite{mass}.  In the last line   of the table, values are given  of the root-mean-square
deviation (RMSD)  of the theory under consideration from the experiment: \beq \label{dQalp_avr}
\langle\delta Q_{\alpha}\rangle_{\rm rms} = \sqrt{\frac 1 N \sum_i \left(Q_{\alpha, i}^{\rm
theor}-Q_{\alpha, i}^{\rm exp}\right)^2}. \eeq

From the $\delta Q_\alpha$ results, we can see that the MMM exceeds in accuracy all the
self-consistent methods used, with its RMSD being smaller by a factor of  1.5--2. However, it is worth
to note that the MMM parameters, those of the liquid-drop model and of the Saxon-Woods shell-model
potential, were fitted in Ref.  \cite{Muntian-1,Muntian-2} to characteristics of heavy deformed nuclei
in  the uranium region,  close to the nuclear map region under consideration. Per contra,  the main
part of the EDFs parameters is universal for all nuclei. Among the self-consistent calculations
presented, the SLy4 EDF results in the highest accuracy. The Fayans EDF has  approximately 10 \% worse
RMSD. However, it is worth reminding that the FaNDF$^0$ parameters were fitted in \cite{Fay5} to
characteristics of the spherical nuclei only of the region between calcium and lead chains. Evidently,
more fine tuning of this EDF parameters is necessary, including heavy deformed nuclei into the fitting
procedure. Accuracy of the SkM* EDF is approximately two times worse comparing to other EDFs used.

\begin{table*}[]
  \caption{Differences in  Eq. (\ref{dQalp})  between theoretical predictions of  $\alpha$-decay energies
 of superheavy nuclei from Table~\ref{table1} and the corresponding experimental data. The mark ``{(\rm syst)}'' in the
last column denotes the data from the systematics in \cite{mass}.}
\label{table2}
\begin{tabular}{c c c c c c c c}
\hline
\hline\noalign{\smallskip}
Nuclei\;&\; FaNDF$^0$(surf) \;&\; FaNDF$^0$(vol) \;&\; SLy4 \;&\; SkM* \;&\; MMM\;&\;data from\\

\hline\noalign{\smallskip}

$^{294 }$118&      -0.532&     -0.500&     -0.446 &    -0.480& 0.300 &\cite{Oganes_2006}\\

$^{290  }$Lv&       0.126 &    -0.009&     -0.636 &     0.084  &0.090 &\cite{Oganes_2006}\\

$^{286  }$Fl&       0.477 &     0.313&     -0.651 &     1.364 &0.490 &\cite{Oganes_2006}\\

$^{282  }$Cn&       1.078 &     0.895&      0.776 &     1.578  &0.570 &\cite{mass}{(\rm syst)} \\

$^{278  }$Ds&       0.902  &    0.655&      0.479 &     1.432 &0.170 &\cite{mass}{(\rm syst)} \\

$^{274  }$Hs&      -0.073  &    0.218&      0.300 &     0.362 &-0.020 &\cite{mass}{(\rm syst)} \\

$^{270  }$Sg&      -0.078  &    0.417 &     0.014 &     0.329 &-0.250 &\cite{mass}{(\rm syst)} \\

$^{266  }$Rf&       0.067  &   -0.234 &    -0.498 &     0.958  &-0.500 &\cite{mass}{(\rm syst)} \\

$^{262  }$No&       0.247  &   -0.040 &    -0.601 &     0.682 &-0.390 &\cite{mass}{(\rm syst)} \\

\hline\noalign{\smallskip}

$^{294 }$117&      -0.975 &    -1.069 &      0.429 &    -0.167 &     0.230& \cite{Z117_new} \\

$^{290 }$115&       0.667 &     0.448 &     -0.307 &     0.342  &     0.200& \cite{Z117_new} \\

$^{286 }$113&      -0.296 &    -0.311 &     -0.998 &     1.089   &     0.190& \cite{Oganes_2013}\\

$^{282  }$Rg&       1.344 &     1.521 &      0.857 &     2.332 &     0.670 &  \cite{Z117_new}\\

$^{278  }$Mt&       0.744 &     0.091 &     0.261  &     0.852 &    -0.040& \cite{Z117_new} \\

$^{274  }$Bh&       -0.013 &     0.217 &    -0.059 &     0.315 &    -0.140 & \cite{Z117_new}\\

$^{270  }$Db&      -0.054  &   -0.492  &   -0.377  &    0.356  &    0.320 &  \cite{Z117_new}\\

$^{266  }$Lr&       0.147  &    1.093  &    0.423  &    0.879   &   -0.780  &\cite{mass}{(\rm syst)} \\

$^{262  }$Md&       0.064  &   -0.069  &    -0.489 &     1.121   & - &\cite{mass}{(\rm syst)} \\

\hline\noalign{\smallskip}

$^{293 }$117&      -0.575 &    -0.678  &    -0.204  &   -0.190 &0.350 & \cite{mass}\\

$^{289}$115&        0.836 &     0.647  &    -0.485  &    0.168 &0.220 & \cite{mass}\\

$^{285}$113&       0.241 &     0.124  &    -0.438   &   1.870 &0.180 & \cite{mass}\\

$^{281  }$Rg&      1.440 &     1.593  &     1.309   &   2.326 &  1.070& \cite{Oganes_2013} \\

$^{277  }$Mt&       0.788 &     0.246  &    0.297  &    0.908 &0.130 &\cite{mass}{(\rm syst)} \\

$^{273  }$Bh&      -0.046 &     0.396  &    0.110  &    0.326 &-0.170 &\cite{mass}{(\rm syst)} \\

$^{269  }$Db&       0.106 &     0.249  &    -0.588 &    -0.185 & -0.320 &\cite{mass}{(\rm syst)} \\

$^{265  }$Lr&      -0.068 &    -0.209  &    -0.022 &     1.273 & -0.610&\cite{mass}{(\rm syst)} \\

$^{261  }$Md&      -0.042 &     0.210  &    -0.350 &     1.280  & - &\cite{mass}{(\rm syst)} \\

\hline\noalign{\smallskip}

$^{291  }$Lv&       0.069  &    0.097  &    -0.333  &   -0.092 &0.020 &\cite{Oganes_2006}\\

$^{287  }$Fl&       0.303  &    0.083  &    -1.030  &    1.368 &0.400 &\cite{Oganes_2006}\\

$^{283  }$Cn&       0.836  &    0.605  &     0.829  &    2.061  &    0.490& \cite{Oganes_2006} \\

$^{279  }$Ds&       1.257  &    1.036  &     0.394  &    1.521  &    0.400&\cite{Oganes_2006}  \\

$^{275  }$Hs&     -0.119  &    0.166  &    0.160 &    0.480   &   -0.030 &\cite{Oganes_2006} \\

$^{271  }$Sg&     0.077   &  -0.555   &   -0.116 &    -0.129   &   0.040 & \cite{Oganes_2006}  \\

$^{267  }$Rf&       0.108  &    1.187  &   0.156 &     0.314 & - &\cite{mass}{(\rm syst)} \\

$^{263  }$No&       0.031  &   -0.241  &   -0.713 &     1.634  & -0.550&\cite{mass}{(\rm syst)} \\

\hline\noalign{\smallskip}

$^{288 }$115&       0.766  &    0.358  &    -1.221  &    0.875 &     0.130 & \cite{Oganes_2005}\\

$^{284 }$113&       0.511  &    0.438  &    0.243   &   1.780 &     0.530& \cite{Oganes_2005} \\

$^{280  }$Rg&       1.338  &    1.404  &    1.153   &   1.847 &     0.900& \cite{Oganes_2005}\\

$^{276  }$Mt&       0.831  &    0.543  &    0.640   &   1.069 &     0.240&  \cite{Oganes_2005}\\

$^{272  }$Bh&      -0.042  &    0.592  &     -0.184 &     0.108 &     0.060&  \cite{Oganes_2005}\\

$^{268  }$Db&       0.338  &   -0.015  &     0.254  &   -0.021 &-0.360 &  \cite{Oganes_2005}\\

$^{264  }$Lr&      -0.057  &   -0.066  &    -0.794  &    1.502 &-0.560 &\cite{mass}{(\rm syst)} \\

$^{260  }$Md&      -0.117  &    0.319  &    -0.021  &    1.249 &  - &\cite{mass}{(\rm syst)} \\

\hline\noalign{\smallskip}

$^{287 }$115&      1.038  &    0.772  &    -0.750  &    1.182  &   0.620& \cite{Oganes_2005}\\

$^{283 }$113&      1.144  &    0.970  &    0.870   &   2.149   &   1.000& \cite{Oganes_2005}  \\

$^{279 }$Rg&     0.124  &    0.237  &      1.119   &   1.737   &   0.710& \cite{Oganes_2005}\\

$^{275 }$Mt&      1.459  &    1.396  &     0.255   &   0.579   &   0.010&  \cite{Oganes_2005}\\

$^{271 }$Bh&      -0.480  &    0.270   &    0.067  &   -0.065 &  -0.420 &\cite{mass}{(\rm syst)} \\

$^{267 }$Db&       0.412  &   -0.233   &   -0.488  &    0.478 & -0.490 &\cite{mass}{(\rm syst)} \\

$^{263 }$Lr&       0.062  &   -0.325   &   -0.701  &    1.064 & -0.420 &\cite{mass}{(\rm syst)} \\

$^{259  }$Md&      -0.142  &    0.627   &  -0.019  &    1.242 &  -     & \cite{mass}{(\rm syst)} \\

\hline\noalign{\smallskip}

$\langle\delta Q_{\alpha}\rangle_{\rm rms}$ = & 0.643 & 0.647 & 0.593  & 1.148  &    0.450 &\\

 \hline
\hline
\end{tabular}
\end{table*}

\section {Lifetimes $T_{\alpha}$ with respect to alpha-decay }

\begin{table*}[]
  \caption{The  ${\rm log}_{10}T_{\alpha}$ for superheavy nuclei found with the PS formula
  \cite{Par-Sobich}, Eq. (\ref{Talp}). The meaning of the upper labels
  for experimental data is the same as in Table~\ref{table1}. The asterisks  indicate three cases, where the
  decay is mixed and the total lifetime is given. The percent number shows the weight of
  the $\alpha$-transition.}
\label{table3}
\begin{tabular}{c c c  c c c c c}
\hline \hline \noalign{\smallskip}Nuclei\;&\; FaNDF$^0$(surf) \;&\; FaNDF$^0$(vol) \;&\; SLy4 \;&\; SkM* \;&
 \; MMM \;& ${\rm log}_{10} T_{\alpha}(Q_{\alpha}^{\rm exp})$  \;&\;  exp \\

\hline\noalign{\smallskip}

$^{294}$118 &   -1.52  & -1.60 &  -1.73  &   -1.65  &-3.41   & -2.75 (0.13)    & -3.23 - (-2.71) $\,^{a)}$ \\
$^{290}$Lv&     -1.73  & -1 40 &   0.20  &   -1.63 & -1.64  & -1.43 (0.23)    & -2.27 - (-1.99) $\,^{a)}$ \\
$^{286 }$Fl&   -1.68   & -1 28 &   1.32  &   -3.73 & -1.72  & -0.47 (0.16)    & -0.96 - (-0.77) $\,^{a*)}$, [50\%] \\
\hline\noalign{\smallskip}

$^{294 }$117&    1.67   &  1.94  &  -1.98 &    -0.52 &  -1.50 & -0.94 (0.10) & -1.51 - (-0.84) $\,^{b)}$\\
$^{290 }$115&   -1.34   & -0.79  &  1.25  &   -0.52&  -0.15 &  0.39 (0.11) & -0.10 - 0.56  $\,^{b)}$\\
$^{286 }$113&    2.51   &  0.26  &  4.82  &   -1.37 &   1.05 &  1.61 (0.15) &  0.62 - 0.91  $\,^{b)}$\\
    &              &        &        &        &        &              &  0.60 - 1.11  $\,^{c)}$\\
$^{282  }$Rg&   -1.09   & -1.54  &  0.23  &     -3.49 &   0.76 &  2.80 (0.10) &  2.06 - 2.72  $\,^{b)}$\\
$^{278  }$Mt&   -1.24   &  0.57  &  0.08  &     -1.52 &   0.96 &  0.84 (0.09) & 0.34 - 1.00 $\,^{b)}$\\
$^{274  }$Bh&    2.08   &  1.36  &  2.23  &      1.06  &   2.50 &  2.04 (0.10) & 1.26 - 1.92 $\,^{b)}$\\
$^{270  }$Db&    4.79   &  6.52  &  6.05  &      3.29  &     -  &  4.58 (0.12) & 3.33 - 4.02 $\,^{b)}$\\

\hline\noalign{\smallskip}

$^{293 }$117&    0.15   &   0.42 &  -0.81 &      -0.84&  -2.15 & -1.31 (0.12)  & -1.41 - (-1.02)$\,^{d)}$ \\
$^{289 }$115&   -2.33   &  -1.88 &  1.07  &     -0.69 &  -0.83 & -0.26 (0.13)  & -0.31 - 0.18 $\,^{d)}$   \\
$^{285 }$113&   -0.23   &   0.09 &  1.69  &     -4.14 &  -0.07 &  0.43 (0.14)  & 0.69  - 1.08$\,^{c)}$ \\

\hline\noalign{\smallskip}

$^{291}$Lv&   -0.924 &  -0.995 &  0.12  &     -0.52  &  -0.801 & -0.75 (0.18)  & -1.92  - (-1.40) $\,^{a)}$\\
$^{287}$Fl&   -0.269 &   0.325 &  3.65  &     -2.89 &  -0.524 &  0.55 (0.17)  & -0.43  - (-0.19) $\,^{a)}$\\
$^{283}$Cn&   -1.02  &  -0.409 & -1.00  &     -3.93 &  -0.098 &  1.29 (0.18)  &  0.49  -  0.70   $\,^{a)}$ \\
$^{279}$Ds&   -3.09  &  -2.57  & -0.95  &     -3.70 &  -0.962 &  0.13 (0.18)  &  -0.80 - (-0.60)  $\,^{a*)}$, [10\%]  \\
$^{275}$Hs&    0.955 &   0.120 &  0.14  &     -0.76 &   0.690 &  0.60 (0.18)  &  -0.92 - (-0.39)  $\,^{a)}$\\
$^{271}$Sg&    2.03  &   4.20  &  2.67  &      2.71 &  2.15  &  2.29 (0.17)  &  1.89 - 2.41      $\,^{a*)}$, [70\%]\\

\hline\noalign{\smallskip}

$^{288}$115&   -1.97  &  -0.97  & 3.53       -2.23  &  -0.93   & -0.04 (0.16)  & -1.24 - (-0.72) $\,^{e)}$  \\
$^{284}$113&   -0.81  &  -0.62  & -0.10      -3.83 &  -0.86   &  0.57 (0.17)  & -0.51 - 0.03    $\,^{e)}$\\
$^{280}$Rg&   -2.79  &  -2.94   & -2.34       -3.95 &  -1.72   &  0.70 (0.17)  &  0.36 - 0.90    $\,^{e)}$\\
$^{276}$Mt&   -2.13  &  -1.39   & -1.64       -2.71 &  -0.58   &  0.08 (0.17)  &  -0.33 - 0.20   $\,^{e)}$ \\
$^{272}$Bh&    2.02  &   0.09   &  2.48        1.54 &   1.69   &  1.88 (0.19)  &   0.80 - 1.33   $\,^{e)}$\\

 \hline\noalign{\smallskip}
 $\langle \delta \lg T\rangle_{\rm rms}\;$=&  1.52  &  1.53   & 1.89  &    2.46  &    0.70   &    0.33 & \\
 \hline
\hline
\end{tabular}
\end{table*}

For completeness we recite the commonly used  classical formula by Viola and  Seaborg \cite{VS}. In
particular, it was used in \cite{Oganes_2006}--\cite{Oganes_2012} to connect values of $Q_{\alpha}$
and $T_{\alpha}$ found experimentally. It reads:
 \beq \lg{T_{\alpha}}(Z,N) = (a Z+b) Q_{\alpha}^{-1/2} + (c Z + d) +
h_i, \label{TalpVS} \eeq where $a, b, c, d, h_i$ are adjusted  parameters. Three of them, $h_i,
i=p,n,pn$, are introduced to reproduce a change  of the $\alpha$-decay lifetime  in odd-proton,
odd-neutron and odd-odd nuclei with respect to ``favored'' decays of even-even nuclei with zero
orbital moment $l$ of the $\alpha$-particle. The ground states of mother and daughter odd or
odd-nuclei have usually  different $J^{\pi}$ values, therefore the $\alpha$-transition between them
will be of unfavored type with $l > 0$,  with additional hindrance due to penetration through the
centrifugal barrier. On the other hand, a favored transition may occur to the excited state of the
daughter nucleus.

\begin{table}[b!]
  \caption{The coefficients \cite{Royer} of  the RZ formula (\ref{Talp_Royer}).}
\label{table4}
\begin{tabular}{c c c c}
\hline \hline\noalign{\smallskip}

    & a   &  b  &  c \\

\hline\noalign{\smallskip}

 ee   & -25.31   & -1.1629   & 1.5864  \\

 eo   & -26.65   & -1.0859   & 1.5848  \\

 oe   & -25.68   & -1.1423   & 1.592  \\

 oo   & -29.48   & -1.113    & 1.6971  \\

\hline \hline
\end{tabular}
\end{table}

An optimum set of the parameters of Eq. (\ref{TalpVS}) for uranium and trans-uranium regions can be
found in Ref.  \cite{Par-Sobich}, where  authors modified this formula to five-parameter PS form: \beq
{\rm log}_{10}{T_{\alpha}}(Z,N) = a Z \left[Q_{\alpha}(Z,N) -
 \bar{E}_i\right]^{-1/2} + b Z + c, \label{Talp}\eeq
with the following set of parameters $a = 1.5372$, $b = -0.1607$ and $c = -36.573$. The parameter
$\bar{E}_i$ in (\ref{Talp}) has the meaning of the average  excitation energy of the daughter nucleus,
being zero in the case of even-even nuclei. For other types of nuclei, the following values of
parameters were found in \cite{Par-Sobich}: \bea
\bar{E}_i= \bar{E}_p = 0.113\; \mbox{MeV} & \mbox{for odd-proton}, \nonumber  \\
\bar{E}_i= \bar{E}_n = 0.171\; \mbox{MeV} & \mbox{for odd-neutron}, \nonumber \\
\bar{E}_i= \bar{E}_p + \bar{E}_n & \mbox{for odd-odd nuclei} \label{Talp1}\eea nuclei.  These rather
simple and transparent PS formulas were used in \cite{alpha-13} and they are also used in the present
work. The results are given in  Table~\ref{table3}. The meaning of superscripts for experimental
values is the same as in Table~\ref{table1}. For three parent nuclei, namely  $^{286 }$Fl, $^{279
}$Ds, and $^{271}$Sg, there is a strong competition between $\alpha$-decay and fission. In these
cases, we gave the total lifetime values only and the $\alpha$-decay percentage only, as it was given
in the original experimental works. Recently, there has been several theoretical analysis of such
competition, see e.g. \cite{Qian} and references therein.

\begin{table*}[t!]
  \caption{The ${\rm log}_{10}T_{\alpha}$ values for superheavy nuclei found with the RZ formula
 \cite{Royer}, Eq. (\ref{Talp_Royer}). The meaning of the upper labels
  for the experimental data is the same as in tables 1 and 3.}
\label{table5}
\begin{tabular}{c c c  c c c c c}
\hline \hline\noalign{\smallskip}
Nuclei\;&\; FaNDF$^0$(surf) \;&\; FaNDF$^0$(vol) \;&\; SLy4 \;&\; SkM* \;&
 \; MMM \;& ${\rm log}_{10} T_{\alpha}(Q_{\alpha}^{\rm exp})$  \;&\;  exp \\

\hline\noalign{\smallskip}

$^{294}$118&   -2.14 &  -2.22 &  -2.35  &     -2.27 &  -4.09 & -3.41 (0.14)  & -3.23 - (-2.71) $\,^{a)}$  \\
$^{290}$Lv&   -2.34 &  -2.00 &   -0.34  &     -2.23 &  -2.25 & -2.02 (0.20)  & -2.27 - (-1.99) $\,^{a)}$ \\
$^{286}$Fl&   -2.27 &  -1.85 &    0.83  &     -4.39 &  -2.30 & -1.02 (0.16)  & -0.96 - (-0.77) $\,^{a*)}$, [50\%]\\

 \hline\noalign{\smallskip}

$^{294}$117&    1.57 &   1.86 & -2.30 &    -0.75 &  -1.79  & -1.19 (0.11)  & -1.51 - (-0.84) $\,^{b)}$ \\
$^{290}$115&   -1.65 &  -1.07 &  1.09 &     -0.78 &  -0.38  &  0.19 (0.11)  & -0.10 - 0.56    $\,^{b)}$ \\
$^{286}$113&    2.39 &   2.44 &  4.83 &     -1.71 &   0.86  &  1.44 (0.16)  &  0.62 - 0.91  $\,^{b)}$ \\
        &              &        &        &        &        &             &  0.60 - 1.11  $\,^{c)}$\\
$^{282}$Rg&   -1.44 &  -1.92 &  -0.05  &     -3.99 &   0.51  &  2.67 (0.11)  &  2.06 - 2.72  $\,^{b)}$ \\
$^{278}$Mt&   -1.62 &   0.29 &  -0.23  &     -1.92 &   0.69  &  0.57 (0.09)  & 0.34 - 1.00 $\,^{b)}$ \\
$^{274}$Bh&    1.85 &   1.09 &   2.01  &      0.77 &   2.29  &  1.81 (0.10)  & 1.26 - 1.92 $\,^{b)}$ \\
$^{270}$Db&    4.66 &   6.47 &   5.98  &      3.10 &   3.23  &  4.45 (0.12)  & 3.33 - 4.02 $\,^{b)}$ \\

\hline\noalign{\smallskip}
$^{293}$117&   -0.33 &  -0.05 &  -1.30 &      -1.34 &  -2.67 & -1.82 (0.13)  & -1.41 - (-1.02)$\,^{d)}$ \\
$^{289}$115&   -2.85 &  -2.39 &   0.62 &      -1.18 &  -1.31 & -0.73 (0.13)  & -0.31 - 0.18 $\,^{d)}$ \\
$^{285}$113&   -0.70 &  -0.38 &   1.26 &      -4.68 &  -0.53 & -0.03 (0.14)  & 0.69  - 1.08$\,^{c)}$  \\

\hline\noalign{\smallskip}
$^{291}$Lv&   -1.22 &  -1.30 &  -0.18  &     -0.81 &  -1.10 & -1.05 (0.18)  & -1.92 - (-1.40) $\,^{a)}$\\
$^{287}$Fl&   -0.57 &   0.02 &   3.36  &     -3.22 &  -0.83 &  0.25 (0.17)  & -0.43  - (-0.19) $\,^{a)}$ \\
$^{283}$Cn&   -1.33 &  -0.72 &  -1.32  &     -4.27 &  -0.41 &  0.98 (0.18)  & 0.49  -  0.70   $\,^{a)}$  \\
$^{279}$Ds&   -3.43 &  -2.90 &  -1.27  &     -4.04 &  -1.29 & -0.19 (0.17)  & -0.80 - (-0.60) $\,^{a*)}$, [10\%]  \\
$^{275}$Hs&    0.63 &  -0.20 &  -0.19  &     -1.08 &   0.37 &  0.28 (0.18)  & -0.92 - (-0.39)  $\,^{a)}$ \\
$^{271}$Sg&    1.71 &  3.88  &   2.35  &      2.39 &   1.83 &  1.96 (0.27)  & 1.89 - 2.41      $\,^{a*)}$, [70\%] \\

 \hline\noalign{\smallskip}

$^{288}$115&   -2.29 &  -1.22 &  3.54  &     -2.56 &  -1.17 & -0.60 (0.17)  &  -1.24 - (-0.72) $\,^{e)}$  \\
$^{284}$113&   -1.08 &  -0.88 & -0.33  &     -4.29 &  -1.13 &  0.38 (0.18)  &  -0.51 - 0.03    $\,^{e)}$ \\
$^{280}$Rg&   -3.20 &  -3.37 &  -2.73  &     -4.44 &  -2.07 &  0.49 (0.18)  &   0.36 - 0.90    $\,^{e)}$ \\
$^{276}$Mt&   -2.53 &  -1.75 &  -2.02  &     -3.15 &  -0.89 & -0.19 (0.18)  &  -0.33 - 0.20   $\,^{e)}$   \\
$^{272}$Bh&    1.82 &  -0.21 &   2.30  &      1.32 &   1.48 &  1.68 (0.20)  &   0.80 - 1.33   $\,^{e)}$  \\

 \hline\noalign{\smallskip}
$\langle \delta \lg T\rangle_{\rm rms}\;$=&  1.67 &  1.64  &   1.87  &    2.82  &    0.87 &  0.23              & \\
 \hline
\hline
\end{tabular}
\end{table*}

Since the formula (\ref{Talp1}) for $T_{\alpha}$ is essentially empirical, it is reasonable to examine
some alternative for it. The empirical  formula for $T_{\alpha}$ of favored $\alpha$-transitions by
Royer \cite{Royer1}, or its modification by Royer and Zhang  \cite{Royer}, has the form of
 \beq {\rm log}_{10}{T_{\alpha}}(Z,N) = a + b A^{1/6}
+ \frac {cZ} {\sqrt{Q_{\alpha}}}. \label{Talp_Royer}\eeq Here the coefficients $a,\,b,\,c$ are
different for four different kinds of nuclei, see Table~\ref{table4}. The  abbreviation `eo' means
even $Z$, odd $N$, and so on.

The unfavored $\alpha$-transitions occur in odd and odd-odd nuclei, provided that  the $J^{\pi}$
values of the parent and daughter nuclei do not coincide, resulting  in the $\alpha$-particle orbital
moment $l$ being nonzero. Several  generalizations of Eq. (\ref{Talp_Royer}), which take  into account
additional hindrance due to the contribution of the centrifugal barrier, have been  presented
\cite{Denisov-1,Denisov-2,Royer2,Dong}. Here, we use a simple parameter free  ansatz for this
$l$-dependent addendum by Dong {\it et al.} \cite{Dong}: \beq {\rm log}_{10}{T_{\alpha}}(l) = {\rm
log}_{10}{T_{\alpha}}(l=0) + \frac {l(l+1)} {\sqrt{(A-4)(Z-2)A^{-2/3}}}. \label{Talp_Dong}\eeq For the
first term of this formula the RZ recipe, Eq. (\ref{Talp_Royer}), was used.

Recently Wang {\it et al.} \cite{Wang} used a modification of Eq. (\ref{Talp_Dong}), with four
additional parameters for the $l$-dependent term. In this work, the role of the centrifugal term was
examined in detail. In the sample of 341 $\alpha$-transitions, with  $J^{\pi}$ values known  from
\cite{Denisov-1} for both the parent and daughter nuclei, the average difference \beq \langle \delta
\lg T\rangle_{\rm rms}  = \sqrt{\sum_i \left({\rm log}_{10}T_{\alpha}^{\rm theor} - {\rm
log}_{10}T_{\alpha}^{\rm exp}\right)^2/N} \label{delT_rms}\eeq between the theoretical predictions
(with the experimental values of $Q_{\alpha}$ used) and the experimental values for the lifetimes were
found in different models. At first, the initial Royer formula (\ref{Talp_Royer}) was used and then,
the one (\ref{Talp_Dong}) with the $l$-dependent  term included. It turned out that the gain for the
due to the $\langle \delta \lg T\rangle_{\rm rms}$ value is about 0.1. To be more exact, from $\langle
\delta \lg T\rangle_{\rm rms}=0.587$ without the $l$-term to  $\langle \delta \lg T\rangle_{\rm
rms}=0.481$ with it. Of course, this gain is different for different kinds of nuclei. It is equal to
zero for even-even nuclei, being about 0.2 for odd-odd ones. The use of the modified centrifugal term
in \cite{Wang} results in a rather small additional gain, $\langle \delta \lg T\rangle_{\rm
rms}=0.433$. The corresponding values of $\langle \delta \lg T\rangle_{\rm rms}=0.536$ in
\cite{Denisov-2} and $\langle \delta \lg T\rangle_{\rm rms}=0.561$ in \cite{Royer2} are also given in
\cite{Wang}. The predictions for lifetimes found in this work with the use of different theoretical
values of $Q_{\alpha}$ are essentially worse. For example, for the Skyrme HFB-27 EDF the average
deviation from the experiment $\langle \delta \lg T\rangle_{\rm rms}=1.617$ was found. The values of
$\langle \delta \lg T\rangle_{\rm rms}$ found with Eq. (\ref{delT_rms}) for each theoretical column in
Table~\ref{table3} are given in the last line of this table. In this calculation, we excluded three
cases of mixed decays.

In contrast to \cite{Wang}, the present work addresses odd and odd-odd superheavy nuclei for  which
the characteristics are often  not known from the  experiment. Therefore, it is reasonable to use for
systematic calculations of the ${\rm log}_{10} T_{\alpha}(Q_{\alpha}^{\rm exp})$ values  Eq.
(\ref{Talp_Dong}) assuming that all the $\alpha$-transitions are favored, i.e. putting $l=0$. Thus, we
use, in fact, the RZ formula (\ref{Talp_Royer}). In Table~\ref{table5}, we present the results of such
calculations for the same set of  nuclei as in Table~\ref{table3}. In both of  the tables, we found
from Eq. (\ref{delT_rms}) for each kind of the theory used the average difference between the
theoretical prediction for ${\rm log}_{10}T_{\alpha}$ and the corresponding experimental value,
 three cases of mixed decays again being excluded from the averaging procedure.
A comparison of the values of $\langle \delta \lg T\rangle_{\rm rms}$ in the columns of ${\rm
log}_{10} T_{\alpha}(Q_{\alpha}^{\rm exp})$ in tables III and V is a direct comparison of the accuracy
of Eq. (\ref{Talp1}) from \cite{Par-Sobich} and Eq. (\ref{Talp_Royer}) from \cite{Royer}. We see that
the latter is a bit more accurate that is not strange as it contains twelve fitted parameters in
comparison with five ones in the first case.

We can now compare the accuracy of the different theoretical methods in the description of the
$\alpha$-decay lifetimes. We see that both the methods to calculate the lifetimes lead to
approximately the same accuracy, the Parkhomenko--Sobiczewski method turning  out to be a bit better.
Evidently, the accuracy of the RZ method could be made higher if all the $\alpha$-transitions were not
considered as favored.  Again, the MMM approach is more accurate than all the self-consistent methods
used. For this characteristic, the Fayans method with the FaNDF$^0$ EDF and both models for pairing
exceed a bit in accuracy the SLy4 EDF and significantly the SkM* one. Note that the $\langle \delta
\lg T\rangle_{\rm rms}$ value  for the Fayans EDF is approximately the same as that found in
\cite{Wang} for the HFB-27 EDF, the most accurate member of the family in predicting nuclear masses
Skyrme EDFs.

In order to estimate the role of the $l$-dependent term in Eq. (\ref{Talp_Dong}) we chose three
$\alpha$-decays in the chain of $^{293}117$ for which we obtained ${\rm log}_{10}
T_{\alpha}(Q_{\alpha}^{\rm exp}) < {\rm log}_{10} T_{\alpha}^{\rm exp}$. Due to hindrance effect of
the  centrifugal barrier, its inclusion may improve the agreement with the data. The calculated
results are given in Table~\ref{table6}. We see that, indeed, the inclusion of the $l$-dependent term
does improve the situation with the use of Eq. (\ref{Talp_Dong}). The values of ${\rm log}_{10}
T_{\alpha}(Q_{\alpha}^{\rm exp})$ agree with the data at $l=3, l=4$ for the $^{293}117$ and
$^{289}115$ parent nuclei and at $l=4, l=5$ for the $^{285}113$ one. Unfortunately, this remedy can
help only in the cases where the inequality ${\rm log}_{10} T_{\alpha}(Q_{\alpha}^{\rm exp}) < {\rm
log}_{10} T_{\alpha}^{\rm exp}$ holds, whereas the opposite sign of this inequality takes place in
many cases in Table~\ref{table5}. For example, this happens for the $^{275}108$ nucleus for which we
obtained ${\rm log}_{10} T_{\alpha}(Q_{\alpha}^{\rm exp}) - {\rm log}_{10} T_{\alpha}^{\rm exp} =
0.487$. However, as the analysis in \cite{Wang} showed the average accuracy of Eq. (\ref{Talp_Dong})
and analogous formulas in \cite{Wang} for ${\rm log}_{10} T_{\alpha}$ are of the order of 0.5, the
disagreement under discussion is not extraordinary.

\begin{table}[]
  \caption{The role of the $l$-dependent term   in formula (\ref{Talp_Dong}). The meaning of the upper labels
  for the experimental data is the same as in Table~\ref{table1}.}
\label{table6}
\begin{tabular}{c c c c c}
\hline \hline\noalign{\smallskip}

 Nuclei   &  \;\; l \;\; &  ${\rm log}_{10} T_{\alpha}(Q_{\alpha}^{\rm exp})$ &\;  exp\\

\hline\noalign{\smallskip}

 $^{293}117$   &  0 & -1.82  & -1.41 - (-1.02)$\,^{d)}$\\
          &  1 & -1.74  & \\
          &  2 & -1.60  & \\
          &  3 & -1.38  & \\
          &  4 & -1.09  & \\
          &  5 & -0.72  & \\

\hline\noalign{\smallskip}
 $^{289}115$   &  0 & -0.73 & -0.31 - 0.18 $\,^{d)}$\\
          &  1 & -0.66 & \\
          &  2 & -0.51 & \\
          &  3 & -0.29 & \\
          &  4 &  0.01 & \\
          &  5 &  0.37 & \\

\hline\noalign{\smallskip}

 $^{285}113$   &  0 & -0.03  & 0.69  - 1.08$\,^{c)}$ \\
          &  1 &  0.05  & \\
          &  2 &  0.20   & \\
          &  3 &  0.42  & \\
          &  4 &  0.72  & \\
          &  5 &  1.09  & \\

\hline \hline
\end{tabular}
\end{table}

\section {Conclusion}
Alpha-decay energies $Q_{\alpha}$ for several chains of super-heavy nuclei are found within the SCMF
approach by employing the Fayans functional FaNDF$^0$. Two models for the effective pairing force were
used, the volume and the surface pairing. The results are compared to the experimental data and
predictions of two Skyrme functionals, SLy4 \cite{sly4to7} and SkM* \cite{skms}. Predictions of the
macro-micro method \cite{Muntian-1,Muntian-2} are also considered. The Fayans EDF results in the
average deviation from experimental energies by $\langle\delta Q_{\alpha}\rangle_{\rm rms}^{\rm
tot}=0.643\;$MeV  with the surface pairing and 0.647 MeV with  the volume pairing. These values are
slightly larger  than the corresponding SLy4 value of 0.593 MeV but significantly less than the SkM*
value of 1.148 MeV. However, in this problem all the considered self-consistent methods give in the
MMM, the corresponding value being only 0.454 MeV. It is worth stressing that the FaNDF$^0$ EDF used
here was found in \cite{Fay5}  by adjusting to  masses and radii of spherical nuclei only, from the
calcium to the lead region. Therefore a readjustment of its parameters with the use of all the nuclear
chart is desirable. As a first step in this direction, we plan to adjust spin-orbit and effective
tensor constants similarly to that made in \cite{DF3-a} for spherical nuclei, which turned out to be
successful. As a second step, the optimization of the EDF parameters at deformed HFB level, by
utilizing input data on heavy nuclei, similarly to \cite{Kor10}, should be performed.

The corresponding $\alpha$-decay lifetimes are calculated with the  use of the semi-phenomenological
five-parameter  PS formula  \cite{Par-Sobich}, or, alternatively, the twelve-parameter RZ one
\cite{Royer}. The role of the $l$-dependent term for the unfavored $\alpha$-transitions in the form of
\cite{Dong} was also examined. The accuracy of the method itself to calculate the lifetimes can be
estimated by comparing the ${\rm log}_{10} T_{\alpha}(Q_{\alpha}^{\rm exp})$ values with the
experimental data of  ${\rm log}_{10} T_{\alpha}$. For the bulk of 24 $\alpha-$transitions between
super-heavy nuclei we examined, the RZ formula gives an  average deviation of $\langle \delta \lg
T\rangle_{\rm rms}=0.23$, whereas it is equal to 0.33 for the PS case. Thus, the first method looks
more accurate itself. However, the accuracy of all calculations with the use of theoretical
$Q_{\alpha}$ values turns out a bit worse in the RZ case.

Any inaccuracy in finding $Q_{\alpha}$ leads to a defect in reproducing the $T_{\alpha}$ values, the
scale of disagreement being even enhanced. Again, the MMM exceeds in accuracy all the self-consistent
calculations. The corresponding value of $\langle \delta \lg T\rangle_{\rm rms}$ is equal to 0.69 with
the use of the PS formula and to 0.87, with the RZ one. In this problem the Fayans approach has some
advantage over the other self-consistent methods used. For this, we obtained $\langle \delta \lg
T\rangle_{\rm rms} \simeq 1.50$ with the PS formula and $\simeq 1.65$ for the RZ one. For comparison,
the corresponding values are 1.89 and 1.87 for the SLy4 EDF, and also 2.46 and 2.82 for the SkM* one.
Note that the average deviation from experiment for the Skyrme HFB-27 EDF found in \cite{Wang} is
$\langle \delta \lg T\rangle_{\rm rms}\simeq 1.5$, i.e. it is comparable to that for the Fayans EDF.

In conclusion, we should again stress that readjustment  of the Fayans EDF parameters with the use of
wider bulk of nuclei, including the deformed and heavy ones, is necessary to attempt to reach the
level of accuracy in reproducing the $Q_{\alpha}$ and $T_{\alpha}$ values comparable  to that of the
MMM.

\section{Acknowledgment}
This work was supported with Grants  Nos. 16-12-10155 and   16-12-10161 from Russian Science
Foundation. It was also partly supported  by the RFBR Grant 16-02-00228. This work was also supported
(M.K.) by the Academy of Finland under the Centre of Excellence Programme 2012--2017 (Nuclear and
Accelerator Based Physics Programme at JYFL) and FIDIPRO programme. Calculations are partially made on
the Computer Center of NRC ``KI''.

{}

\end{document}